\documentclass[fleqn,twoside]{article}
\usepackage{espcrc2}

\usepackage{graphicx}
\usepackage[figuresright]{rotating}

\newcommand{\AmS}{{\protect\the\textfont2
  A\kern-.1667em\lower.5ex\hbox{M}\kern-.125emS}}

\title{Matrix Model and Ginsparg-Wilson Relation}

\author{Keiichi Nagao\address{The Institute of Particle and Nuclear
        Studies, \\
High Energy Accelerator Research Organization (KEK), Tsukuba 305-0801, Japan}%
        \thanks{This talk is mainly based on the work with 
H.Aoki (Saga Univ.) and S.Iso (KEK).}
}
       
\begin{document}

\begin{abstract}
We discuss that the Ginsparg-Wilson relation, 
which has the key role in the recent development of 
constructing lattice chiral gauge theory, can play an important 
role to define chiral structures in finite matrix models and 
noncommutative geometries.

\vspace{1pc}
\end{abstract}

\maketitle

\section{Introduction}

Various matrix models have been proposed toward 
the nonperturbative formulations of the superstring theory. 
IKKT matrix model\cite{IKKT} is one of the proposals, 
and its several properties have been investigated\cite{IKKTreview}.   
In the matrix model space-time is described in matrices, so 
noncommutative (NC) geometry\cite{Connes} naturally appears 
in itself. 
The realization of chiral fermion 
is one of the interesting issues from the standpoint of 
the unification of space-time and matter since 
the chiral structure of fermions may 
play important roles in quantization of space-time.
 
One way to realize the $4$ dimensional chiral 
fermions in the matrix model is the Kaluza-Klein 
compactification with the non-trivial index. 
The construction of the non-trivial index in finite NC geometry 
or matrix model is thus an important subject. 
Topologically nontrivial configurations in finite NC 
geometries have been constructed based on algebraic 
K-theory and projective modules in much literature, but 
it would be better if we could find other prescriptions 
which are more natural in physical point of view.
One of the possibilities is the utilization of 
the admissibility condition\cite{Luscher:1981zq} and the 
Ginsparg-Wilson relation\cite{GinspargWilson} 
in lattice gauge theory (LGT). 

In LGT there occurred much development 
recently in the construction of chiral gauge theory. 
The first important observation was 
that, in the presence of the mass defect which is introduced 
as a scalar background in higher $4+x$ dimensions, a chiral 
fermion appears at the defect. 
So far a domain wall fermion ($x=1$)~\cite{dwfermion} 
and a vortex fermion ($x=2$)~\cite{Nagao:2001tc} 
are constructed on the lattice. {}From the former model 
a practical solution to the GW relation\cite{GinspargWilson}, 
the overlap Dirac operator, is obtained~\cite{Neuberger:1998fp}.
The index theorem\cite{Hasenfratzindex,Luscher} and 
the modified chiral symmetry\cite{Luscher,Nieder} are also 
realized, and eventually, an anomaly free abelian chiral gauge 
theory is constructed on the lattice\cite{Luscher:1998du}.

On the other hand in Connes' NC geometry the formulation 
is evolved from the spectral 
triple (${\cal A}$,${\cal H}$,${\cal D}$), where a chirality 
operator and a Dirac operator which anti-commute are 
introduced\cite{Connes}.  
In ref.\cite{AIN2}, we proposed to generalize 
the algebraic relation to GW relation\cite{GinspargWilson} 
so that we can 
define chiral structures in a finite noncommutative geometry.
We will discuss these topics briefly in the following.


\section{GW fermion on finite NC geometry}
First we overview the relevant works shortly.

The first literature related to this context 
is found in ref.\cite{kitsu-nishi}, where 
overlap formula is applied to the unitary IIB matrix model. 
At this time, however, the real importance of the GW relation and 
the relation between them were not known yet.
Next in ref.\cite{balaGW} GW relation on fuzzy 2-sphere 
was discussed for free fermions 
in connection to the fermion doubling problems.
On NC torus the lattice structure is visible\cite{NClattice}, 
so the usual overlap 
Dirac operator\cite{Neuberger:1998fp} can be put naively 
on NC torus as it is.
The authors in ref.\cite{nishimozo} discussed 
the chiral gauge theory on NC torus using the overlap 
Dirac operator.
In the paper\cite{immiyd}  
the gauge field was introduced to the Dirac operator 
in ref.\cite{balaGW}, but the modified 
chiral symmetry was not maintained since the gauge field 
was introduced linearly in the Dirac operator. 
Dirac operator which has the exact chiral symmetry and 
no doubling species for general gauge fields has not been 
realized until the 
appearance of the paper\cite{AIN2}.

The prescription proposed in ref.\cite{AIN2} is as follows.
First we introduce two hermitian chirality operators, $\Gamma$ and 
$\hat\Gamma=\frac{H}{\sqrt{H^2}}$, which satisfy 
$\Gamma^2=\hat{\Gamma}^2=1$.  
$H$ includes the differential operator and the 
gauge field. 
The Dirac operator $D_{GW}$ is defined by 
$1- \Gamma \hat{\Gamma} = f(a,\Gamma) D_{GW}$, 
where $a$ is a small parameter. 
$H$ and the function $f$ must be defined  
so that the $D_{GW}$ has no doublers and 
the correct behavior in the commutative limit ($a \rightarrow 0$). 
$D_{GW}$ satisfies GW relation\cite{GinspargWilson}, 
$\Gamma D_{GW}+D_{GW} \hat{\Gamma}=0$, 
and we have the index theorem: 
${\rm{index}}D_{GW}\equiv (n_+ - n_-)=\frac{1}{2}
{\cal T}r(\Gamma+\hat{\Gamma})$.
Under the modified chiral transformation 
$\delta \Psi =i \lambda \hat{\Gamma} \Psi$, 
$\delta \bar{\Psi} = i \bar{\Psi}\lambda \Gamma$,
the fermionic action $S_F={\rm tr}(\bar\Psi D_{GW} \Psi)$
is invariant, while the Jacobian produces 
$q(\lambda)=
\frac{1}{2}{\cal T}r(\lambda \hat{\Gamma} +\lambda \Gamma)$, 
which is expected to provide the topological charge density and 
the index for $\lambda=1$.

\subsection{Example on fuzzy 2-sphere}
We give the concrete example of $H$ and $f$ for fuzzy 2-sphere case 
setting the NC coordinates as $x_i=\alpha L_i$, where 
$L_i$'s are $2L+1$ dimensional irreducible representation matrices 
of $SU(2)$ algebra. The radius of the sphere is then given by 
$\rho=\alpha\sqrt{L(L+1)}$. 
One example set of $H$ and $f$ are $H=\Gamma^R + a D_w$ and 
$f=-a\Gamma^R$, where $a=1/(L+1/2)$ and $R$ means 
that the operator acts from the right on the matrices. 
$D_w$ and $\Gamma^R$ are given by 
$D_w=\sigma_i({\cal L}_i+\rho a_i)+1$ and 
$\Gamma^R = \frac{1}{2L+1}(2\sigma_i L_i^R -1)$, 
where $i$ runs from $1$ to $3$. 
$a_i$'s are gauge fields and ${\cal L}_i$'s are 
Killing vectors satisfying ${\cal L}_i M =[L_i,M]$ for any 
hermitian matrix $M$. 
We denote here the Dirac operator constructed from these $H$ and 
$f$ as $D_{AIN}$.

We can show that the Chern-character 
is correctly produced from 
the Jacobian in the commutative limit; 
$q(\lambda) = 2\rho \int\frac{d \Omega}{4 \pi} 
\lambda \epsilon_{ijk} x_i \partial_j a'_k$
where $a'_i$ is a tangential component of $a_i$: 
$a'_i = {\epsilon_{ijk}x_j a_k / \rho }$\cite{AIN2}.
 
The properties of $D_{AIN}$ and other types 
$D_{WW}$ and $D_{GKP}$ are summarized in Table 1.
$D_{WW}$, $D_{GKP}$ and their properties related to doublers 
are seen in refs.\cite{Carow-Watamura:1996wg},\cite{Grosse:1994ed} 
and \cite{balagovi,balaGW}, respectively.
Here we note that $D_w$ is equal to $D_{GKP}$ 
and that $D_{WW}$ has no chiral anomaly. 
The source of the chiral anomaly in $D_{GKP}$ is 
the breaking in a cut-off scale of the action under 
the chiral transformation\cite{AIN1}, while 
that in $D_{AIN}$ is the Jacobian\cite{AIN2}.
These facts suggest that some kind of Nielsen-Ninomiya's 
theorem exists in matrix model and NC geometry.

\paragraph{{\bf Non-trivial index on fuzzy 2-sphere}}
Based on the construction in ref.\cite{AIN2}, a concrete 
example of the non-trivial index 
is constructed with the projective module method 
in ref.\cite{balimmi}.
Using the same method 
we can construct the TP monopole on fuzzy 2-sphere\cite{AIN3}. 
However, the index operator is modified and 
the original interpretation is lost.  
This is an open problem.



\begin{table*}[htb]
\caption{The properties of three types of Dirac operators on fuzzy 2-sphere}
\begin{center}
\renewcommand{\arraystretch}{1.2}
\begin{tabular}{|c@{\quad\vrule width0.8pt\quad}l|c|c|c|}
\hline
Dirac operator &\multicolumn{2}{c|}{chiral symmetry} & no doublers & counterpart in LGT \\
\hline
$D_{WW}$ & $D_{WW} \Gamma + \Gamma D_{WW} =0$ & $\bigcirc$ & $\times$ & naive fermion \\
\hline 
$D_{GKP}$ & $D_{GKP} \Gamma + \Gamma D_{GKP} ={\cal O}(1/L)$ & $\times$ & $\bigcirc$ & Wilson fermion \\
\hline  
$D_{AIN}$ & $D_{AIN} \hat\Gamma + \Gamma D_{AIN} =0$ & $\bigcirc$ & $\bigcirc$ & GW fermion \\
\hline 
\end{tabular}
\end{center}
\end{table*}

\subsection{Example on NC torus}

GW fermion on NC torus is defined with the overlap Dirac 
operator\cite{Neuberger:1998fp} in LGT as it is\cite{nishimozo}. 
In addition to the usual chirality operator $\gamma_{d+1}$, 
the other one $\hat\gamma=\frac{H}{\sqrt{H^2}}$ is introduced. 
Here $H = \gamma_{d+1}(m_0-aD_{\rm w} )$ 
with the Wilson-Dirac operator,   
$D_{\rm w}={1\over2}\left[\gamma_\mu(\nabla_\mu^*+\nabla_\mu)
-ar\nabla_\mu^*\nabla_\mu\right]$, 
where $m_0$ and~$r$ are free parameters. 
Using the above chirality operators, the GW Dirac operator 
is expressed as 
$D_{GW}={1\over a}\left[1-\gamma_{d+1} \hat{\gamma} \right]$, 
which is free of species doubling 
if $0<\frac{m_0}{r}<2$\cite{Neuberger:1998fp}. 

The Jacobian is shown to produce the correct chiral anomaly 
in arbitrary dimensions ($d=2n$) using the 
result in ref.\cite{Fujiwara:2002xh}. 
The final expression is written as 
$q(\lambda)=\frac{(-1)^n}{(4\pi)^n n!} 
\epsilon_{\nu_n \mu_n \cdots \nu_1 \mu_1} 
\int d^d x \ \lambda (x) \star F_{\nu_n \mu_n}(x) 
\star \cdots \star F_{\nu_1 \mu_1}(x)$\cite{Iso:2002jc}.

\paragraph{{\bf Non-trivial index on NC torus}}
In LGT the non-trivial topological structure in gauge field space 
is realized by the admissibility 
condition\cite{Luscher:1981zq}, 
which requires that the fluctuation of the gauge field strength 
is sufficiently small. 
On NC torus GW fermion includes the gauge field in 
the compact form, so it is expected that 
the non-trivial topology may be realized with the 
condition 
not appealing to projective modules.
The index can be studied by the spectral flow method. 
The gauge field configurations 
which satisfy the admissibility condition and produce 
non-trivial indices are indeed found 
numerically on NC torus\cite{KNtorus}. 
This result shows that the non-trivial topological 
structure is naturally realized 
in the gauge field space on NC torus.

\section{Discussion}
GW relation has important roles in matrix model as well as in LGT. 
It is desirable to construct many examples of GW fermions 
on various higher-dimensional NC geometries. 
We hope that not only the topology of gauge field space but 
also that of space-time can be classified with 
the index operator or its generalization.

%


\end{document}